# Robust Video Watermarking Schemes in Phase domain Using Binary Phase Shift Keying

K.Meenakshi [1], Ch.Srinivasa Rao [2], and K. Satya Prasad [3]

[1] Department of E.C.E, GRIET, Hyderabad, A.P., India
Email: meenakshi@griet.in

[2.] Department of E.C.E, University College of Engineering JNTUK, Vizianagaram,. A.P, India,
Email: Ch_rao@ rediffmail.com

[3]. Department of E.C.E, University College of Engineering JNTUK, Kakinada, A.P, India.
Email: prasad_kodati@yahoo.co.in

*Abstract*— This paper presents a robust video watermarking scheme in Discrete Fourier Transform (DFT) and Sequency-ordered Complex Hadamard Transform (SCHT). The DFT and SCHT coefficients are complex and consist of both magnitude and phase and are well suited to adopt phase shift keying techniques to embed the watermark. In the proposed schemes, the phases of DFT and SCHT coefficients are modified to convey watermark information using binary phase shift keying in cover video. Low amplitude block selection (LABS) is used to improve transparency, amplitude boost to improve the resistance of watermark from signal processing and compression attacks and spread spectrum technique is used for encrypting watermark in order to protect it from third party. It is observed that both algorithms showing more or less same robustness but SCHT offers high transparency, simple implementation and less computational cost than DFT.

I. INTRODUCTION

Digital media has become an integral part of modern lives. So, with the explosion of internet and wireless networks mass duplication and the unfettered distribution of copy right material become so easy that there is an urgent need for protecting the ownership rights. So digital watermarking has been proposed one solution to the problem of protecting intellectual rights [1-2]. It is a potential method to discourage the tracking of illegal copying and distribution of data.

In watermarking, transparency means when viewed by a third party there must be no distinction between original and watermarked video. Robustness, invisibility and security are three major requirements of any watermarking system. A variety of techniques are proposed to embed robust watermark in images. These techniques are generally classified into two categories-one processed in spatial domain and other processed in frequency domain. In the spatial domain [13, 14] the watermark is embedded directly by altering the pixel intensities .Example, hiding in least significant bit planes of frames of video. The disadvantage of these techniques is the watermark can be easily erased by lossy compression techniques. On the other hand the transform domain technique[1, 12] insert watermark in transform coefficients of original or cover image .The important advantage of frequency domain techniques is they are more resistant to signal processing and compression attacks. Popular transforms are Discrete Fourier Transform (DFT) [2- 4] , Walsh-Hadamard Transform (WHT) [12] , Unified Complex Hadamard Transform (UCHT) [7-8] and Discrete Fractional Randon transform (DFRNT) [1] Though watermarking image and video share some similarities, new problems need to be addressed in video watermarking. [16] They are problems like temporal redundancy in addition to spatial redundancy present in image and attacks unique to video such as frame dropping; frame swapping, frame averaging and MPEG compression Motivated by the good performance of 2D image watermarking techniques proposed in a scene change [15] watermarking approach, a video watermarking based on phase modulation of DFT and SCHT coefficients is proposed in this paper. The application of video frames in DFT/SCHT can be viewed as 3D DFT/SCHT with two dimensions in space and one dimension in time. These algorithms are blind, and they do not require original video for extraction.

According to communication theory the phase is more immune to noise compared to amplitude and frequency. The approach used in watermarking of phase domain is the host image is segmented into $8 \times 8$ blocks and each block is mapped into the transform domain. Only the portions which are most significant to the image integrity are watermarked. In Ref. [5], amplitude boost (AB) and the low amplitude block selection (LABS) methods are used to enhance further the robustness of watermarking scheme using DFT. The scheme resistant to most of the common signal processing attacks, but it does not withstand to the phase perturbation attack [17]. Similar technique is used for embedding watermark using Sequency Complex Hadamard

Transform [14, 15, 16]. The advantage of SCHT is it can withstand phase perturbation attack. So far all these methods are applied to still images. The same techniques are extended to video for achieving high transparency and robustness. The kernel of DFT contain sines and cosines whereas SCHT contain kernel coefficients as {+1,-1, i, -i}. Out of all the complex transforms, SCHT is the lowest computational transform. So SCHT is more suitable for real time implementation than DFT.

II. WATERMARKING EMBEDDINGALGORITHMS

The design and implementation DFT and SCHT based watermarking algorithm for test video sequences is presented in this paper. A robust watermarking must preserve the transparency and at the same time must be resistant to signal processing and compression attacks. The proposed method is described step by step.

Algorithm 1 Watermarking in phase domain using DFT

*1. Scrambling binary logo*: A binary watermark of size $36 \times 44$ is taken. A pseudorandom binary pattern is generated by seed key PN2. The pseudo binary random pattern is exclusive-or with the binary watermark and scrambled binary watermark is obtained as shown in shown in figure1.

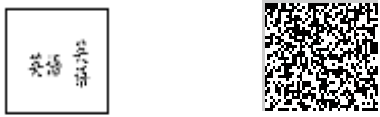

Figure1. $(a)$ Original watermark $(b)$ Permuted watermark.

*2. Spreading the binary sequence:* Unipolar binary sequence (0, 1) is converted to bipolar sequence. It is found bipolar sequence is more resistant to attacks. The bit sequence is spread to three times by sending a bit sequence [1 -1 1] in place of 1 and [-1 1 -1] in place of -1.
The watermark size is 1584 bits. But encrypted information of watermark embedded is 4752 bits. This is to prevent the third party from extracting the original watermark.

*3. RBS*: Rearrange the encrypted watermark from two dimensional arrays into one dimensional array

*4. Watermark Embedding*: For watermarking video, YUV QCIF videos of resolution $144 \times 176$ are used. Divide the video clips into video scenes. Process the frames of video scene. The frames of YUV video are in RGB color standard. The RGB color standard is converted into YUV. The luminance component is less affected by modification than chrominance components (U & V). So extract luminance component for watermark embedding.
For each frame do the following.
a) Each frame is partitioned into non overlapping blocks of 8x8.
b) On each 8x8 block 2D DFT is applied. The coefficients of DFT are magnitude and phase.
c) Low amplitude block selection is used for selecting some of 8x8 blocks for watermark embedding in each frame to preserve transparency.
d) Amplitude boost is used to boost the amplitude of coefficient of DFT in each 8x8 block of every frame

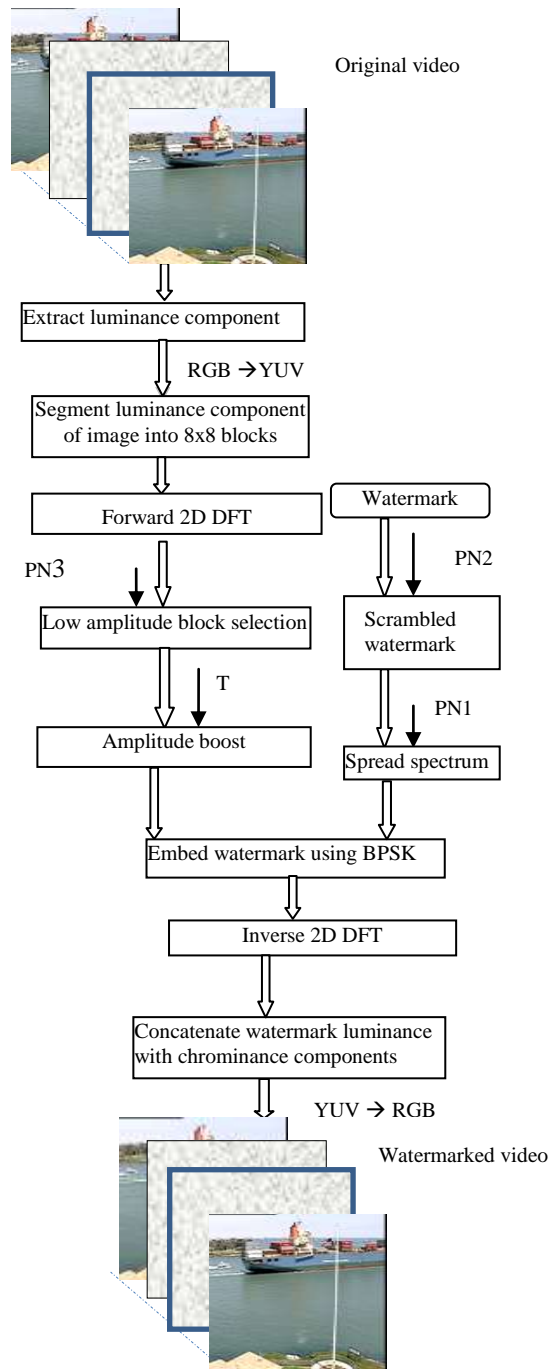

Figure 2. Watermark Embedding Algorithm in DFT based video watermarking.

If it is below certain threshold T so that the phase distortion under attacks can be keeps below a certain level.

e) The Phase of DFT coefficient of selected 8x8 blocks Q (1, 1) is raised from $\phi$ as shown in $\phi'$ according

$$\beta(m) = \begin{cases} \dfrac{\pi}{2} & if \quad m = 0 \\ -\dfrac{\pi}{2} & if \quad m = 1 \end{cases} \quad (1)$$

f) Inverse 2D DFT is performed on the transformed frame to obtain watermarked luminance. Later it is combined with chrominance components U and V to reconstruct watermarked YUV frame.
g) Convert YUV watermarked frame from YUV to RGB color standard.
h) Concatenate all frames to obtain watermarked video.

In DFT based video watermarking the luminance value is not completely real due to round off errors before inverse DFT applied. So in DFT based video the luminance component must be made conjugate symmetric before the inverse DFT is applied.

Algorithm 2: Watermarking video in phase domain using SCHT is described below:

The steps of scrambling logo, spreading the binary sequence, RBS and extracting luminance from video is similar to the algorithm 1.

For each frame do the following.
a) Divide each frame into non overlapping blocks of 8x8.
b) The 2D-SCHT is performed on each 8x8 block to transform the whole image from spatial domain to frequency domain. The 8×8 kernel is given by

$$H_N = \begin{vmatrix} 1 & 1 & 1 & 1 & 1 & 1 & 1 & 1 \\ 1 & 1 & i & i & -1 & -1 & -i & -i \\ 1 & i & -1 & -i & 1 & i & -1 & -i \\ 1 & i & -i & 1 & -1 & -i & i & -1 \\ 1 & -1 & 1 & -1 & 1 & -1 & 1 & -1 \\ 1 & -1 & i & -i & -1 & 1 & -i & i \\ 1 & -i & -1 & i & 1 & -i & -1 & i \\ 1 & -i & -i & -1 & -1 & i & i & 1 \end{vmatrix} \quad (2)$$

c) The 2D-SCHT has been defined in [13, 14]

$$S_N = C_N X_N C_N^T \quad (3)$$

$$C_N = \dfrac{1}{\sqrt{N}} H_N^* \quad (4)$$

where $H_N^*$ is the complex conjugate of SCHT matrix $H_N$, $Z_N$ is the 8×8 image block and $Y_N$ is the 2D SCHT coefficients of dimensions 8×8. The SCHT contains a real image are the complex values which consists of magnitude and phase and Binary phase shift keying can be used to embed the watermark in luminance component of video scenes. Then Low amplitude block Selection, Amplitude boost and phase shift keying is done similar to DFT.
d) Apply inverse 2D SCHT is performed on the transformed frame to obtain watermarked luminance.

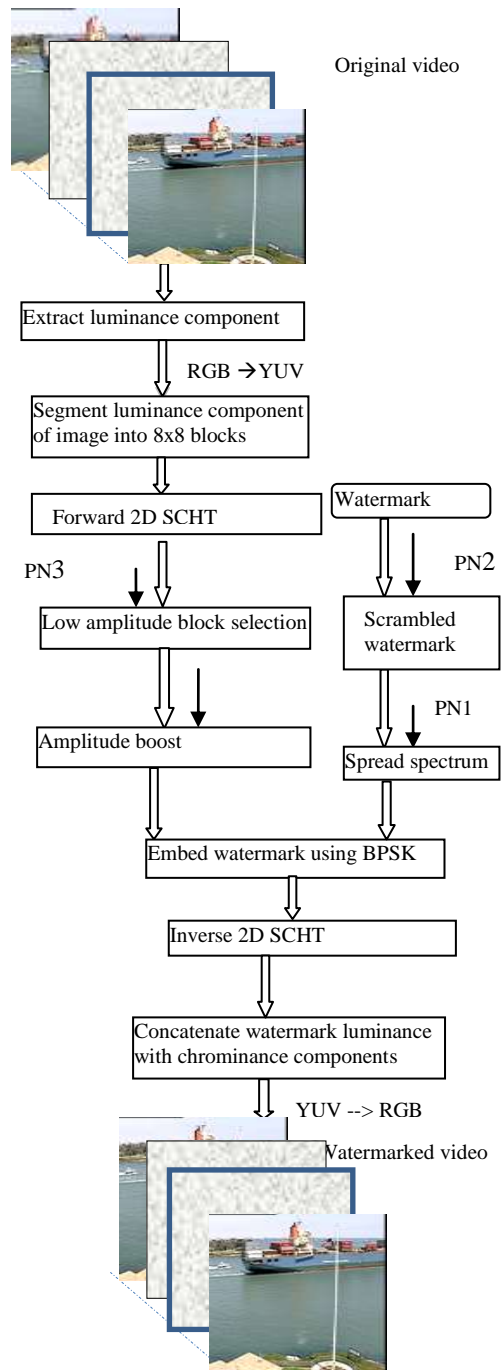

Figure3. Watermark Embedding Algorithm in SCHT based video watermarking.

e) Combining with U and V to reconstruct watermarked YUV frame.
f) Convert YUV watermarked frame from YUV to RGB.
But the luminance of SCHT component is real. So the problem of making conjugate symmetric of DFT does not arise in SCHT.

## Watermark Extraction algorithms

In the extraction, the watermark bits must be extracted from the same position where they are embedded.

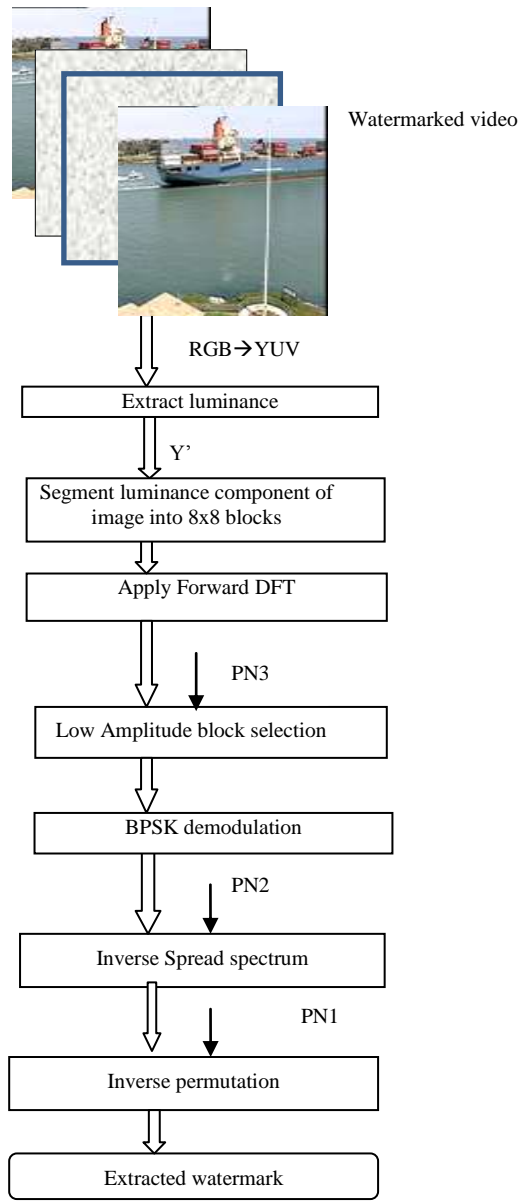

Figure4.Watermark extraction using DFT

1. The input to the extraction process is watermarked video.
2. Watermarked video is segmented into frames.
3. Frames in embedding process are selected using a key.
4. Each frame is divided into non overlapping blocks of $8 \times 8$ and 2D DFT is applied.
5. Select the $8 \times 8$ blocks where watermark is embedded based on Low amplitude block selection.
6. Using PSK demodulation extract the watermark from selected DFT coefficients from the selected blocks in each selected frames.
7. Extract all the secret bits from all frames and form one dimensional array.
8. Convert 1D watermark array into 2d array.
9. Apply inverse spread spectrum and convert binary bit sequence from -1 to 0 and 1 to1.

10. Apply inverse permutation to recover the original watermark.

Algorithm2: The watermark extraction process is similar to DFT except the forward transform used is SCHT.

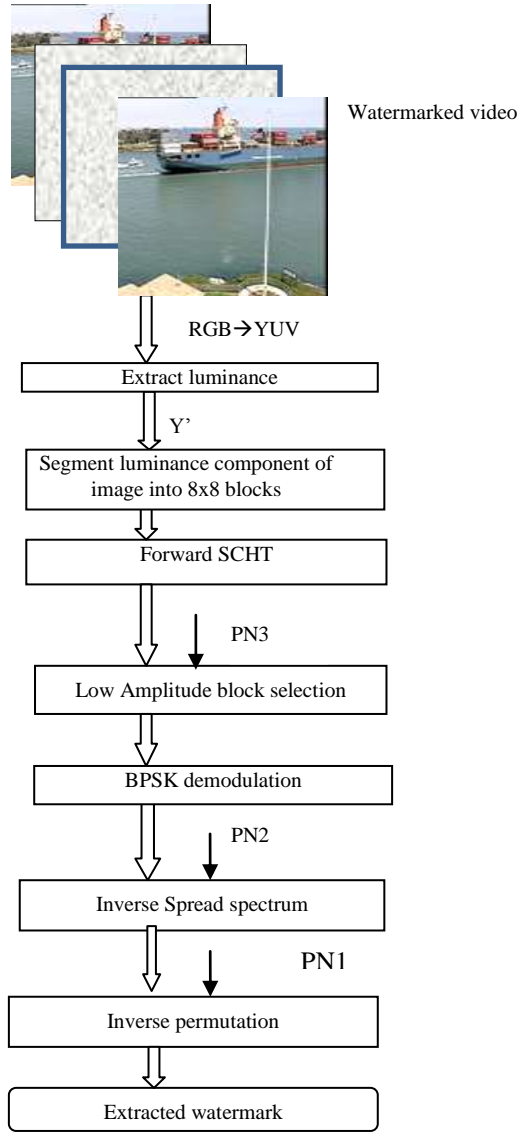

Figure5. Watermark Extraction Algorithm in SCHT based video watermarking.

in the process of SCHT demodulation, the angle $\phi"$ is processed to recover the secret bits embedded from watermarked video. According to minimum distance decision rule, m' is defined as

$$m_R(r) = \begin{cases} 1 & when \quad |\phi" \geq 0|; \\ -1 & when \quad |\phi" \prec 0|; \end{cases} \quad (5)$$

### iv. Experimental results

Imperceptibility is an important factor used in watermarking. PSNR is a measure of transparency.

$$PSNR = 20\log_{10}\left(\frac{Max_i}{\sqrt{MSE}}\right) \qquad (6)$$

Where MSE is mean squared error between the cover frame $F$ and the watermarked frame $\overline{F}$.

$$MSE = \frac{1}{MN}\sum_{i=1}^{M}\sum_{J=1}^{N}\left|F_{ij}-\overline{F}_{ij}\right|^2 \qquad (7)$$

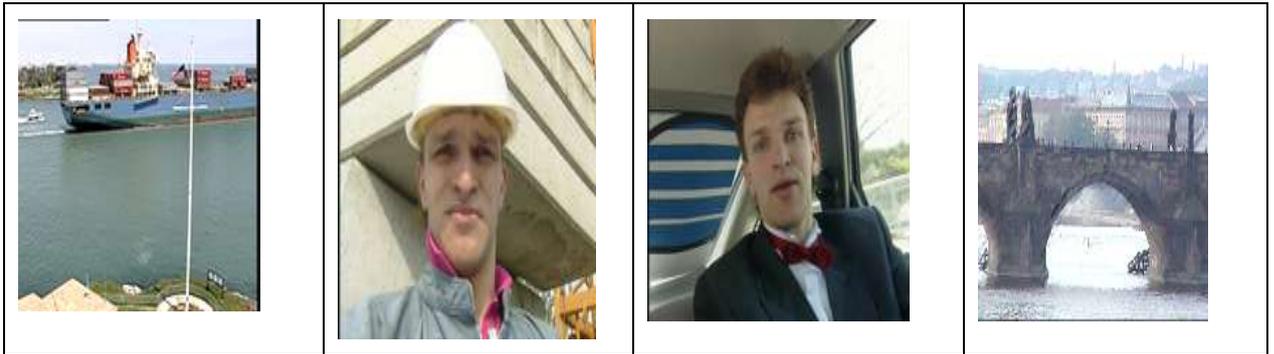

Figure 6. Original videos.

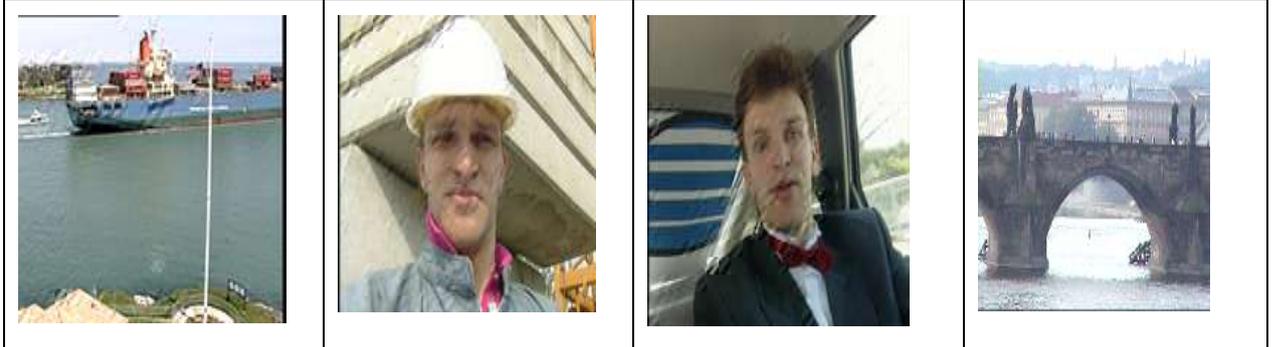

Figure 7 Frames of Watermarked Videos Using DFT

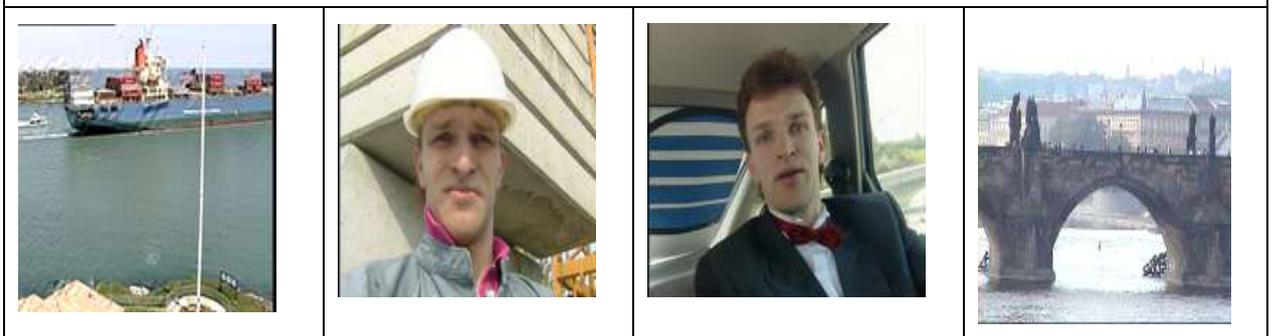

Figure 8. Frames of Watermarked Videos Using SCHT

Table1
Attacks applied to watermarked video

| | | |
|---|---|---|
| 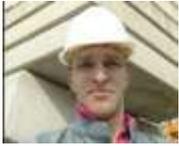 (a) Resizing | 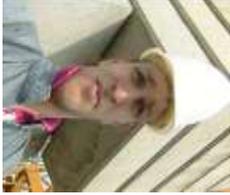 (b) Rotation | 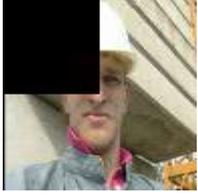 (c) Cropping quarter portion of frame |
| 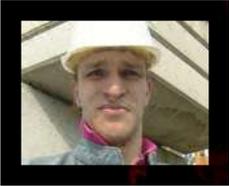 (d) Symmetrical crop | 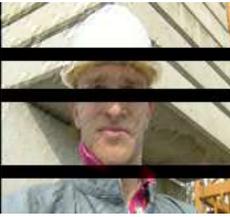 (e) Painting | 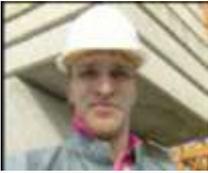 (f) Low pass filtering |
| 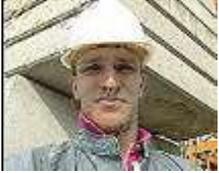 (g) Sharpening | 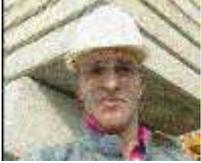 (h) Gaussian noise | 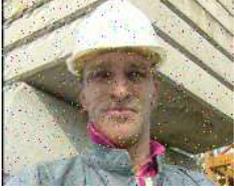 (i) Salt & Pepper noise |
| 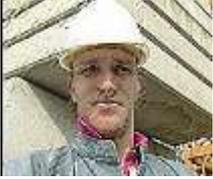 (j) Phase perturbation | 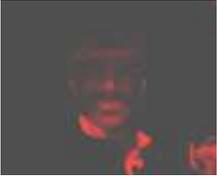 (k) Histogram Equalization | 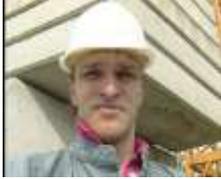 (l) MPEG compression, |

The watermark similarity measure is necessary to provide the judgment of robustness of watermark. Therefore, a qualitative measure used is normalized cross correlation (NC) and BER (Bit error rate)

$$NC = \frac{\sum_i \sum_j w(i,j) w'(i,j)}{\sum_i \sum_j w(i,j)^2} \qquad (8)$$

The test videos taken are the foreman, car phone, and container and Bridge QCIF YUV videos. The logo Used is National Cheng Kung University. The number of bits in watermark is 1584. The number of bits after spread spectrum expansion is 4752 bits. The total number of $8\times 8$ blocks in each video frame is 396. Out of them 264 are selected on LABS strategy. T used in amplitude boost controls both transparency as well as robustness. As T is increased, transparency suffers, robustness improves and the opposite is true if T is decreased. As T is boosted the image is degrades as shown in table 2. The image is degraded slowly in DFT compared to the SCHT. But in SCHT, robustness can be improved by increasing T without degrading quality of image because of its high peak to signal noise ratio.

Table 2.
Peak signal to Noise ratio and Normalized cross correlation under no attack.

| Video | PSNR of DFT based video watermarking | PSNR of SCHT based video watermarking technique. | NC of DFT based video watermarking | NC of proposed video watermarking technique. |
|---|---|---|---|---|
| Car phone | 29.999 | 37.701 | 1 | 1 |
| Bridge | 31.7285 | 43.60 | 1 | 1 |
| Foreman | 26.6549 | 36.96 | 1 | 1 |
| container | 31.1348 | 37.571 | 1 | 1 |

*a)Video frame resizing*: The frame is resized to $72\times 88$ from $144\times 176$ and it is resized back to $144\times 176$. In video watermarking algorithms rescaling is an important attack. Normally rescaling the image in fact introduces high frequency components in the frequency domain while weakening the low and middle frequency components .Since the watermark bits are embedded in the phase of low sequency components , rescaling will lead the degradation of SCHT has high degradation at low threshhold values of amplitude boost compared to DFT

*b)Video rotation*: Video is rotated 90 degrres clockwise followed by anti clockwise direction before decoding.

*c)Video cropping:* One quarter of each frame is cropped before watermark decrypted.

*d)Video Cropping(central portion)* : Cropping on all four sides of video before watermark decoded.

*e)Low pass filtering*: The filter is rotationally symmetric Gaussian filter of the size $(3\times 3)$ with the standard deviation of 1.4.The embedded frame is filtered before proceeding to the detection phase.

*f)Low pass filtering*: The filter is rotationally symmetric Gaussian filter of the size $(3\times 3)$ with the standard deviation of 1.4.The embedded frame is filtered before proceeding to the detection phase.

*g)Sharpening:* The contrast of the image is adjusted to enhance image quality before extracting the watermark

*h)Addition of Gaussian Noise*: Gaussian noise of mean 0 and standard deviation 0.01 is added to watermarked video. For addition of Gaussian noise, SCHT is showing superior performance over DFT at low threshhold values used in amplitude boost.

*i)Addition of Salt &Pepper Noise:* Salt &Pepper Noise is added to watermarked video.

*j)Phase pertubaation*:First , normally distributed noise with mean $\frac{\pi}{4}$ and variance 0.01 is generated

*k)Histogram Equilization* :watermarked video is subjected to Histogram Equilization.

*l) MPEG compression*: watermarked video is subjected to MPEG compression of different quality factors.from 20 to 100.The SCHT showed superior performance over DFT for MPEG compression as shown in Table 7.

The proposed algorithm is subjected to various attacks and extracted quality of the watermark suggests that the proposed method is superior and robust against attacks. But it is found that the quality of extracted watermark depends on threshold used in amplitude boost. For attacks like painting, quarter cropping, symmetrical cropping, the BER is decreased with increasing T so under no attack T=10 and 13 taken for DFT and SCHT. But during attacks T is taken 22 in both cases.

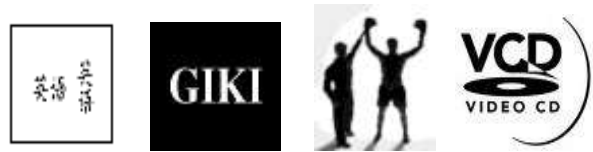

Watermark1   Watermark2   Watermark3   Watermark4

Fig.9. Different watermarks used in the watermarked videos

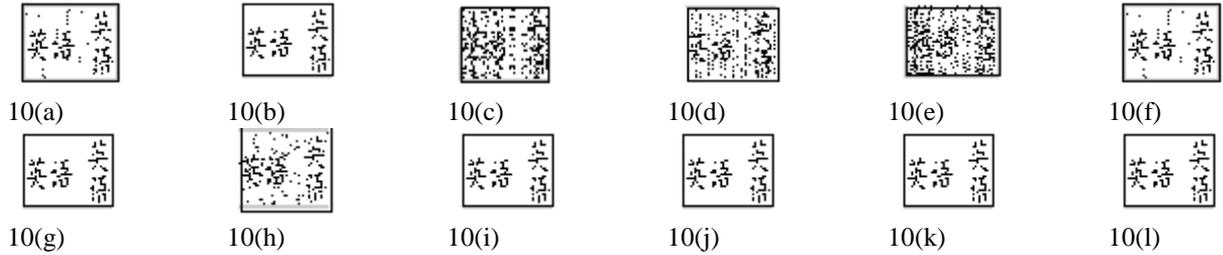

10(a)　　　10(b)　　　10(c)　　　10(d)　　　10(e)　　　10(f)

10(g)　　　10(h)　　　10(i)　　　10(j)　　　10(k)　　　10(l)

Fig (10) The corresponding extracted watermarks in SCHT based video watermarking: a The resized, b the rotated, c the cropped (a quarter image), d the cropped (centralportion remained), e the painted, f the filtered, g the sharpened, h Gaussian noise, and i salt &pepper noise k. phase perturbation attacked. l)MPEG compression

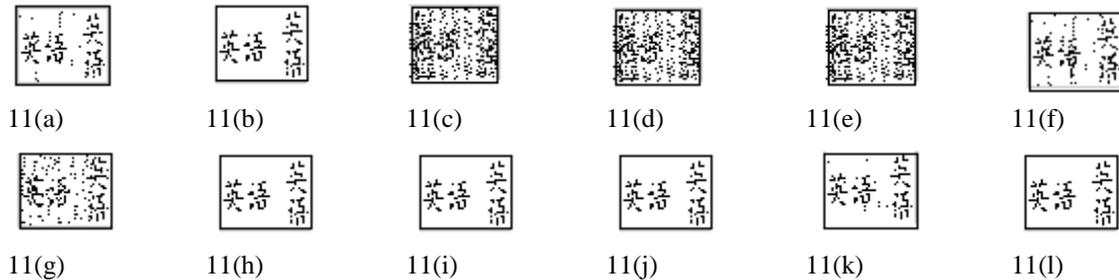

11(a)　　　11(b)　　　11(c)　　　11(d)　　　11(e)　　　11(f)

11(g)　　　11(h)　　　11(i)　　　11(j)　　　11(k)　　　11(l)

Fig (11) The corresponding extracted watermarks in DFT based video watermarking: a The resized, b the rotated, c the cropped (a quarter image), d the cropped (centralportion remained), e the painted, f the filtered, g the sharpened, h Gaussian noise, and i salt &pepper noise k. phase perturbation attacked. l)MPEG compression

Table 3
B.E.R and NC of extracted watermark

| Attack type | Parameters | SCHT eBits | SCHT NCC | DFT eBits | DFT NCC |
|---|---|---|---|---|---|
| a) Scaling | $144 \times 176$ to $72 \times 88$ | 93 | 0.91 | 116 | 0.89 |
| b) Rotation | 90 degrees rotated clockwise | 0 | 1 | 0 | 1 |
| c) Cropping1 | Cropping ¼ | 295 | 0.8311 | 295 | 0.8311 |
| d) Cropping 11 | Central portion remained | 331 | 0.7910 | 193 | 0.8782 |
| e) Painting | Painting 3 bars. | 193 | 0.8782 | 184 | 0.8838 |
| f) Low pass filtering | Size $3 \times 3$, $\sigma = 1.4$ | 0 | 0 | 0 | 1 |
| g) sharpening | Size $3 \times 3$ | 0 | 1 | 0 | 1 |
| h) Gaussian noise | Mean=0 and standard deviation:0.01 | 89 | 0.945 | 33 | 0.96 |
| i) Salt &pepper noise |  | 0 | 1 | 0 | 1 |
| j) Phase pertubation | Mean=$\frac{\pi}{4}$ and var=0.01 | 0 | 1 | 0 | 1 |
| k) Histogram Equilization |  | 5 | 1 | 0 | 1 |
| l) MPEG compression | 100% | 0 | 1 | 0 | 1 |

*m) Frame dropping:* Any video sequence may contain a large number of redundancies between the frames. So, the frame drooping attackis very common and effective on video watermarking. The watermark is embedded into the frames of a scene, and due to the large amount of redundancies between frames, the watermark is robust still frames dropped upto 60%.Different percentages of the video frames are dropped and then Normalised cross correlation is calculated between original and extracted.

*n)Frame averaging* is another common attack in video watermarking. The attackers can use multiple frames and try to eliminate the watermarks by statastical averaging. In the proposed algorithm different watermarks are embedded in each scene change. Even watermark is frame averaged in one scene, the watermark can be extracted from another scene.

*o) Frame swapping*: It is another common attack used even if frames are swapped the watermark must be extracted. The watermark is extracted even if the frames swapped up to 60%

The proposed algorithm is subjected to various attacks and extracted quality of the watermark suggests that the proposed method is superior and robust against attacks. But it is found that the quality of extracted watermark depends on threshold used in amplitude boost. For attacks like painting, quarter cropping, symmetrical cropping, the BER is decreased with increasing T so under no attack T=10 and 13 taken for DFT and SCHT. But during attacks T is taken 22 in both cases.

It has been found that the peak signal to noise ratio of SCHT is superior to DFT. The blocking artifacts in DFT is completely absent in SCHT based watermarked video. Though the results of DFT and SCHT are comparable, the DFT is computationally complex than SCHT. The DFT requires 56 Complex additions and multiplications and SCHT requires 24 complex additions and multiplications for $8\times 8$ block. So for 2 minutes video sequence there is a computational saving of $4752\times 32\times 60$ complex multiplications and additions with SCHT over DFT.

[17] Aye Aung, Boon Poh Ng, Susanto Rahardja A Robust Watermarking scheme using Sequency ordered complex Hadamard transform in J.sign Process Syst .(2010).

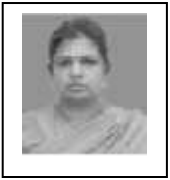
**K.Meenakshi**   received the B.Tech degree from  Nagarjuna University, AP, India in Electronics and Communication Engineering and Masters degree in .Digital systems and Computer electronics from JNTU, Hyderabad, AP ,India  in 2006. She is currently pursuing her Ph.D in JNTU, Kakinda.Her areas of    interest are digital image processing and Embedded systems
.

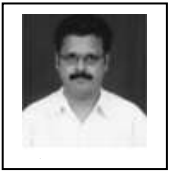
**Ch. Srinivasa rao** is currently working as Professor of ECE and Head of the Department at JNTUK University College of Engineering, Vizianagaram, AP,  India. He obtained his Ph.D. in Digital Image Processing area from University College of Engineering, JNTUK, Kakinada, A.P, India. He received M. Tech degree from the same institute. He published 20 Research papers in reputed International Journals and Conferences.

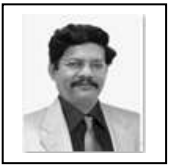
**K. Satya Prasad**: Received his Ph.D. degree from IIT Madras, India. He is presently working as professor in ECE department, JNTU college of Engineering Kakinada. . He has more than 30 years of teaching and research experience. He published 30 research papers in international and 20 research papers in National journals. His area of interests is digital signal and image processing, communications, adhoc networks etc.